\documentclass[twoside,twocolumn,english,aps,showpacs,prl,superscriptaddress,longbibliography]{revtex4-2}
\usepackage[T1]{fontenc}
\usepackage{color}
\usepackage{bm}
\usepackage{amsmath}
\usepackage{amsthm}
\usepackage{amssymb}
\usepackage{graphicx}
\usepackage{subfigure}
\usepackage{esint}
\usepackage{mathrsfs}
\usepackage{booktabs}
\usepackage{multirow}
\usepackage{makecell}
\usepackage{verbatim}
\usepackage{physics}
\usepackage[plainpages = false, pdfpagelabels, bookmarksnumbered = true,
                 breaklinks = true,
                 linktocpage,
                 colorlinks = true,
                 linkcolor = blue,
                 urlcolor  = blue,
                 citecolor = blue,
                 anchorcolor = green,
                 hyperindex = true,
                 hyperfigures
                 ]{hyperref} 
\usepackage{dcolumn}  

\begin{document}

\preprint{APS/123-QED}

\title{Charge Density Wave Driven Topological Phase Transition in Vortices}
\author{Zhenhua Zhu}
\affiliation{State Key Laboratory of Low Dimensional Quantum Physics, Department of Physics, Tsinghua University, Beijing, 100084, China}
\affiliation{Frontier Science Center for Quantum Information, Beijing 100184, China}
\author{Ziqiang Wang}
\email{Corresponding to: ziqiang.wang@bc.edu}
\affiliation{Department of Physics, Boston College, Chestnut Hill, Massachusetts, USA}
\author{Dong E. Liu}
\email{Corresponding to: dongeliu@mail.tsinghua.edu.cn}
\affiliation{State Key Laboratory of Low Dimensional Quantum Physics, Department of Physics, Tsinghua University, Beijing, 100084, China}
\affiliation{Frontier Science Center for Quantum Information, Beijing 100184, China}
\affiliation{Beijing Academy of Quantum Information Sciences, Beijing 100193, China}
\affiliation{Hefei National Laboratory, Hefei 230088, China}

\begin{abstract}
   The interplay between charge density waves (CDWs) and superconductivity is a central theme in quantum materials, yet how CDW phase textures govern vortex topology remains poorly understood. We develop a theoretical framework showing that the phase of a stripe CDW can switch a magnetic vortex between topological and trivial regimes. Motivated by recent experiments, we propose two candidate mechanisms enabling phase-controlled switching of vortex topology. In a direct-modulation scenario, the CDW acts as a periodic potential that locally renormalizes band parameters and can induce topological transitions, but it generally cannot reproduce the symmetric node/antinode trend without fine tuning. In contrast, in an inversion-symmetry-breaking (ISB) scenario, a CDW node pinned to the vortex center breaks local inversion and
   allows for the mixture of spin-triplet pairing of Cooper pairs,
   producing a robust topological transition when this component dominates. Our results suggests CDW phase as a possible local handle to tune and test vortex topology.
\end{abstract}

\maketitle

\section{Introduction}
In type-II superconductors (SC), magnetic flux penetrates the sample as quantized vortices where the superconducting order parameter, $\Delta_0$, is suppressed~\cite{ginzburg2009theory,abrikosov1957magnetic}. This suppression creates a unique environment within vortex cores: it promotes competing orders, such as charge density waves (CDW)~\cite{lake2002antiferromagnetic,hoffman2002four,wu2013emergence,katano2000enhancement,chang2008tuning,edkins2019magnetic}, while simultaneously hosting discrete Caroli–de Gennes–Matricon (CdGM) bound states~\cite{caroli1964bound,berthod2017observation,chen2018discrete,liu2018robust,deng2020bound}. When the SC is intrinsically topological or proximity-coupled to a topological insulator (TI), these bound states may manifest as Majorana zero modes (MZMs)~\cite{fu2008superconducting,hosur2011majorana}, non-Abelian excitations essential for topological quantum computation~\cite{read2000paired,kitaev2001unpaired,nayak2008non}. Indeed, following the observation of topological surface state in FeTeSe~\cite{zhang2018observation}, zero-bias conductance peaks consistent with MZMs have been widely reported in iron-based superconductors~\cite{wang2018evidence,machida2019zero,zhang2021observation,kong2021majorana,li2022ordered}.

\begin{table*}[t]
    \centering
    \caption{Summary of CDW mechanisms, their dimensional effects, and topological properties.}
    \renewcommand{\arraystretch}{1.5} 
    \begin{tabular}{l c l c l c l c}
        \toprule
        \toprule
        \textbf{\parbox{2.2cm}{\centering Mechanism}} & \textbf{Dim} &
        \textbf{\parbox{2.2cm}{\centering Phase sensitivity?}} & 
        \textbf{\parbox{3.0cm}{\centering Matches node/antinode symmetry?}} & 
        \textbf{\parbox{1.5cm}{\centering 
        Tenable in 2D limit?}} & 
        \textbf{\parbox{2.0cm}{Needs fine tuning?}} & 
        \textbf{\parbox{3.0cm}{\centering Discrete vortex MZM expected?}} & \textbf{\parbox{2.0cm}{\centering Status}} \\
        \midrule
        
        \parbox{2.2cm}{\centering Direct modulation} & 3D &
        \parbox{2.2cm}{\centering Yes} & 
        \parbox{3.0cm}{\centering No}
         &\parbox{1.5cm}{\centering No} 
        & 
        \parbox{2.0cm}{\centering Yes} & \parbox{3.0cm}{\centering Not robust} & \parbox{2.0cm}{\centering Disfavored} \\
        \cmidrule{2-8}
         & 2D &
        \parbox{2.2cm}{\centering Weak/unphysical} & 
        \parbox{3.0cm}{\centering No}
         &\parbox{1.5cm}{\centering No} 
        & 
        \parbox{2.0cm}{\centering /} & \parbox{3.0cm}{\centering No} & \parbox{2.0cm}{\centering Discarded} \\
        \midrule
        
        \parbox{2.2cm}{\centering Inversion-symmetry breaking} 
        & 3D &
        \parbox{2.2cm}{\centering Yes} & 
        \parbox{3.0cm}{\centering Yes}
         &\parbox{1.5cm}{\centering No} 
        & 
        \parbox{2.0cm}{\centering No} & \parbox{3.0cm}{\centering No, continuum spectrum} & \parbox{2.0cm}{\centering Discarded for experiment} \\
        \cmidrule{2-8}
         & 2D &
        \parbox{2.2cm}{\centering Yes} & 
        \parbox{3.0cm}{\centering Yes}
         &\parbox{1.5cm}{\centering Yes} 
        & 
        \parbox{2.0cm}{\centering No} & \parbox{3.0cm}{\centering Yes} & \parbox{2.0cm}{\centering Favored mechanism} \\
        \bottomrule
        \bottomrule
    \end{tabular}
    \begin{flushleft}
    \footnotesize{Note: We have assumed the Zeeman coupling sufficiently small compared to the SC gap.}
    \end{flushleft}
    \label{table1}
\end{table*}

While several studies report both topological and conventional vortices in iron-based superconductors~\cite{machida2019zero,kong2021majorana,li2022ordered}, a recent experiment~\cite{liu2026intertwined} revealed a striking and nontrivial correlation: vortex topology is controlled by the relative phase of a coexisting CDW. The vortex core is topological when a CDW node is centered at the core, but trivial when an antinode (CDW maximum or minimum) lies near it. This observation implies a critical coupling between the CDW phase relative to the vortex core and the topological structure of the vortex CdGM spectrum, an ingredient absent from previous theory.

\begin{figure}
    \centering
    \includegraphics[width=0.4\textwidth]{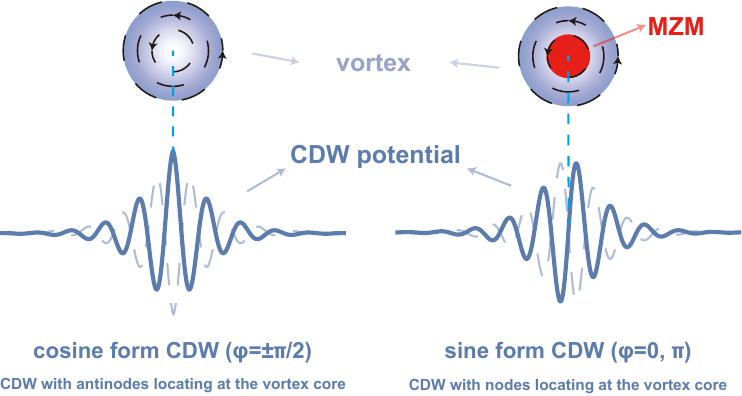}
    \caption{A schematic diagram of 2D vortices in the presence of a CDW. The CDW gradually vanishes outside of the vortex due to the presence of SC. In Ref.~\cite{liu2026intertwined}, MZMs exist when the nodes of CDW locate around the vortex core and disappears when the antinodes of CDW locate there. The dashed line indicated minus-cosine (sine) form CDW potential.}
    \label{setup}
\end{figure}

Motivated by this observation, we develop theoretical models to clarify how a stripe CDW can govern the topology of vortex-bound states shown in Fig.~\ref{setup}.
Our goal is not a material-specific microscopic derivation of the pairing interaction, but a phenomenological identification of the minimal mechanism consistent with the experimentally observed symmetric node/antinode dependence of vortex topology.
As a minimal starting point, we treat the CDW as a periodic perturbation that modulates local band parameters (e.g., the chemical potential and effective mass). In this picture, the modulation can tune the vortex core across a topological–trivial boundary, offering a convenient framework to analyze the topological phase transition of a Fu–Kane vortex.
Beyond this direct modulation mechanism, we further develop a symmetry-based mechanism in which the CDW phase relative to the vortex core changes the local inversion properties of the vortex region. The resulting inversion breaking allows singlet–triplet mixing and can induce a topological effective spin-triplet pairing component~\cite{sato2009topological}.
We discuss each scenario in both 3D and 2D limit. By evaluating these theoretical models to the experiment observations, we identify the 2D inversion-symmetry breaking (ISB) mechanism as the most consistent mechanism of
producing the observed dependence of vortex topology on the CDW phase (summarized in Table.~\ref{table1}).
These results identify the 2D ISB mechanism as the minimal robust explanation for the observed CDW-phase dependence of vortex topology.

\section{Model of a magnetic vortex}
In type-II SC, vortices emerge under an applied magnetic field ($H$) whose magnitude lies between the first and the second critical values, i.e., $H_{c1}<H<H_{c2}$~\cite{ginzburg2009theory,abrikosov1957magnetic}. Here, we assume: 1) the vortex size is sufficiently smaller than the distance between vortices so that the vortices are approximately uncoupled and we can focus on a single vortex~\cite{abrikosov2004nobel,chiu2020scalable}; and 2) CDW is the primary order and provides the leading charge-order effect on vortex topology in the vortex region, while other coexisting orders contribute only subleading corrections.

The three-dimensional (3D) model of a vortex in the iron-based superconductors is~\cite{hosur2011majorana,kawakami2015evolution,rakhmanov2011majorana}:
\begin{equation}\label{3Dmodel}
\begin{split}
H_{3D}=\frac{1}{2}\int dr^3\Psi_{\vec{r}}^\dag \begin{pmatrix}
H_{TI} & \hat{\Delta}(\vec{r})\\
\hat{\Delta}^\dag(\vec{r}) & -H_{TI}
\end{pmatrix}\Psi_{\vec{r}}
\end{split},
\end{equation}
where the TI Hamiltonian is defined as:
\begin{equation}
\begin{split}
H_{TI}=&v_F\tau_x\vec{\sigma}\cdot \vec{p}+[m-\epsilon p^2]\tau_z-E_F+H_{CDW},
\end{split}
\end{equation}
with $\tau$, $\sigma$ the Pauli operators in orbit and spin basis, respectively. The basis is chosen as $\Psi_{\vec{r}}^\dag=(c_{\vec{r}\uparrow1}^\dag,~c_{\vec{r}\uparrow2}^\dag,~c_{\vec{r}\downarrow1}^\dag,~c_{\vec{r}\downarrow2}^\dag,~c_{\vec{r}\downarrow1},~c_{\vec{r}\downarrow2},~-c_{\vec{r}\uparrow1},~-c_{\vec{r}\uparrow2})$ where $c_{\vec{r}sa}^\dag$ creates an electron with the orbit index $a$ and spin index $s$ at the site $\vec{r}$ in 3D space. In this model, $\hat{\Delta}(\vec{r})=\Delta_0\tanh(r/\xi)e^{i\theta}\tau_0\sigma_0$ represents the SC order parameters inside a vortex, where $\xi$ is the coherence length of SC and $(r,~\theta)$ denote the (radial distance, azimuthal angle) in cylindrical coordinates. The TI Hamiltonian $H_{TI}$ contains spin-orbit coupling (SOC) with strength $v_F$, bulk band with gap $2m$ and effective mass parameter $\epsilon$, chemical potential $E_F$, and the mean field Hamiltonian of CDW $H_{CDW}$ which will be discussed later. In the absence of CDW, the system is in the strong TI regime when topological conditions $m\epsilon>0$ and $|E_F|<E_{C}\equiv v_F\sqrt{m/\epsilon}$ are satisfied~\cite{hosur2011majorana}. In this regime, the vortex has a integer-quantized energy spectrum, $E_\mu=\mu\Delta_0^2/E_F$ with $\mu$ an integer, and will host MZM at the surfaces.

Because MZMs persist in the extremely thin limit observed in Ref.~\cite{liu2026intertwined}, a two-dimensional (2D) model warrants detailed investigation. We introduce the 2D BdG Hamiltonian as~\cite{hu2016theory,deng2020bound,chiu2020scalable}:
\begin{equation}\label{2Dmodel}
\begin{split}
H_{2D}=\frac{1}{2}\int dr^2\Psi_{\vec{r}}^\dag \begin{pmatrix}
H_2 & \hat{\Delta}(\vec{r})\\
\hat{\Delta}^\dag(\vec{r}) & -H_2
\end{pmatrix}\Psi_{\vec{r}}
\end{split},
\end{equation}
where the basis is chosen as $\Psi_{\vec{r}}^\dag=(c_{\vec{r}\uparrow}^\dag,~c_{\vec{r}\downarrow}^\dag,~c_{\vec{r}\downarrow},~-c_{\vec{r}\uparrow})$. The 2D band Hamiltonian reads:
\begin{equation}
\begin{split}
H_2=&H_{bulk}+v_F\vec{\sigma}\cdot \vec{p}-E_F+H_{CDW}.
\end{split}
\end{equation}
Here, the SC order parameter becomes $\hat{\Delta}(\vec{r})=\Delta_0 \tanh(r/\xi)e^{i\theta}\sigma_0$ and $(r,~\theta)$ denote the (radial distance, azimuthal angle) of the polar coordinates. Specifically, the $H_{bulk}=0$ case also recovers the effective 2D description of a 3D topological model in the absence of a CDW.

\section{Direct CDW modulation}
As a natural starting point, we first study the topological phase transition (TPT) related to CDW-induced modulation. In the presence of time-reversal symmetry, we usually consider two types of CDW: the onsite type describing the oscillation of electron occupation onsite (oCDW), and the hopping type describing the real nearest-neighbor bond modulations (hCDW, also called as charge bond order)~\cite{feng2021chiral,banerjee2022charge,zeng2023chiral,lin2024impact,banerjee2025charge},
\begin{align}
    H_{oCDW}(\vec{r})&=V_of(\vec{r})\sin(Qx+\varphi_o)c_{\vec{r}}^\dag c_{\vec{r}},\\
    H_{hCDW}(\vec{r})&=2\sum_{\delta \vec{r}}V_hf(\vec{r})\sin(Qx+\varphi_h)c_{\vec{r}}^\dag c_{\vec{r}+\delta\vec{r}},
\end{align}
where $V_o$ and $V_h$ are the corresponding strengths, $f(\vec{r})$ is the real-space decay function, $Q$ is the CDW wave number, $\varphi_{o/h}$ is the CDW phase relative to the vortex center, and $\delta\vec{r}$ is the nearest-neighbor vector.

A strong CDW potential would break rotational symmetry and thereby produce non integer quantization of the CdGM states, contrary to experimental observations~\cite{liu2026intertwined}; we therefore restrict to the weak CDW regime. In this limit, $H_{oCDW}$ and $H_{hCDW}$  act as perturbations, and the leading energy shift of the $\mu$th vortex state is
\begin{equation}
\delta E_\mu^{(1)} \propto \int dr^3 H_{o(h)CDW} \Psi_\mu^*\Psi_\mu,
\end{equation} 
with $\Psi_\mu$ the $\mu$-th eigenstate of $H_{3D/2D}$. Because the vortex region is inversion-symmetric and $|\Psi_\mu|^2$ is inversion-even, $\delta E_\mu^{(1)}$ vanishes for inversion-odd CDW perturbations. Moreover, near the center a sine-form CDW vanishes, whereas a cosine-form CDW remains finite, so the correction depends on the relative phase $\varphi$.

\begin{figure}[t]
\centering
\includegraphics[width=0.5\textwidth]{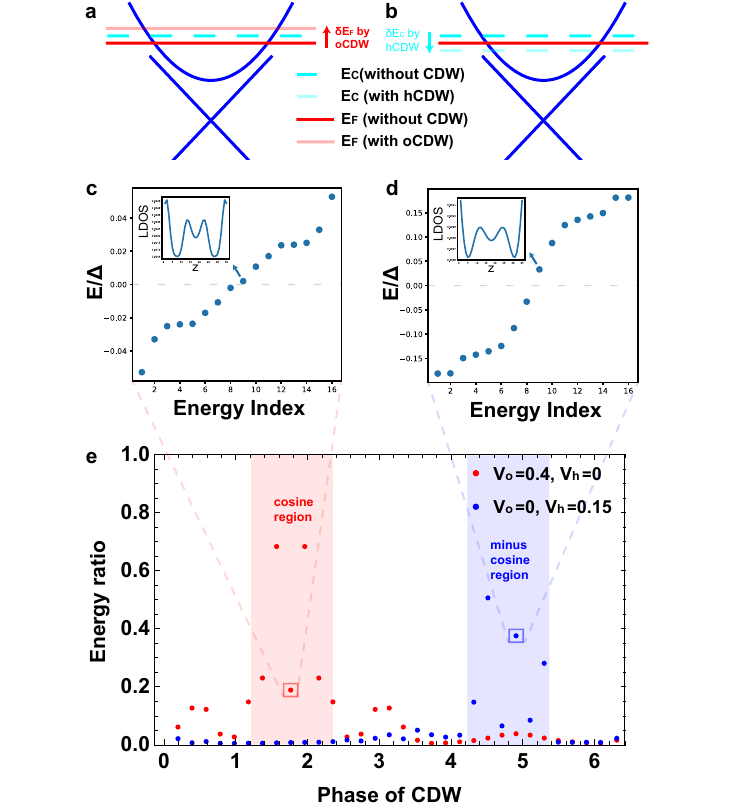}
\caption{\textbf{(a)-(b)}: Schematic diagram of breaking topological conditions in the 3D model by CDW. TPT can occur in cosine (minus-cosine) region by oCDW (hCDW). However, it is difficult to induce TPT in both regions simultaneously since $E_F$ and $E_C$ both shift. \textbf{(c)-(d)}: Energy spectrum of a vortex corresponding to the case in (a) and (b), respectively. The insets are the local density of states (LDOS) at the vortex core along z direction of the state with the lowest energy, which indicates that these states are not surface states ($z=0$ and $z=34$ represent the top and bottom surface of the vortex). We have set $V_o=0.4$, $V_h=0$, $\varphi=\pi/2$ in (c) and $V_o=0$, $V_h=0.15$, $\varphi=3\pi/2$ in (d), respectively. \textbf{(e)}: Energy ratio of the lowest to the first excited vortex bound state as a function of the CDW phase $\varphi$. A large ratio indicates the removal of the MZM and the emergence of a trivial gap. The highlighted regions near $\varphi=\pi/2$ (cosine region) and $\varphi=3\pi/2$ (minus-cosine region) correspond to topologically trivial phases driven by oCDW and hCDW modulations, respectively. In our calculation, the system is a cylinder with radius $R=10$ and height $H=35$. Parameters are: $E_F=0.3$, $\Delta=0.1$, $v_F=0.5$, $\xi=6$, $Q=\pi/2$, $m=0.14$ and $\epsilon=0.25$. In the calculation, we have assumed that $f(\vec{r})=1-\tanh(r/\xi)$. The calculation is performed with the kwant package~\cite{groth2014kwant}.}
\label{V1V2}
\end{figure}

In the 3D model, the vortex topology depends on the band parameters and can therefore be modulated by the CDW. To isolate the resulting TPT, we vary oCDW and hCDW separately, starting from a topological vortex ($E_F \lesssim E_C$, the other cases follow similar argument). With only oCDW ($V_o\neq 0$), the local chemical potential is modulated, and for $\varphi \approx \pi/2$ (cosine region) the effective $E_F$ can exceed $E_C$, driving the vortex trivial. With only hCDW ($V_h\neq 0$), the modulation renormalizes the band parameters $m$ and $\epsilon$ in Eq.~\eqref{3Dmodel}, lowering $E_C$; thus at fixed $E_F$ the condition $E_F>E_C$ can be met near $\varphi \approx 3\pi/2 \bmod 2\pi$ (minus-cosine region), producing a TPT. This phase dependence is summarized in Fig.~\hyperref[V1V2]{2(e)} using the energy ratio between the lowest and first excited vortex bound states: a large ratio indicates a topologically trivial vortex.

When both CDW coexist, $E_F$ and $E_C$ shift simultaneously, making their relative change the decisive factor for the topology. We set the phases of the two CDW components to best match the experiment~\cite{liu2026intertwined} here (see more discussions in \ref{appb}), capturing the essential physics. For a cosine modulation, let oCDW and hCDW induce $\delta E_F=\delta_1$ and $\delta\epsilon=\delta_2$, which indicates the minus-cosine modulation $\delta' E_F\simeq-\delta_1$ and $\delta'\epsilon\simeq-\delta_2$. Demanding a TPT in both scenarios leads to the contradictory requirements: (see details in \ref{appc}):
\begin{equation}\label{TPTcondition}
E_F<v_F\sqrt{m/\epsilon},\qquad
E_F>v_F\sqrt{m/\epsilon}+\mathcal{O}(V^2),
\end{equation}
with $\mathcal{O}(V^2)$ the higher-order correction. Thus, cosine and minus-cosine modulations drive $E_F$ and $E_C$ in opposite directions, and a generic 3D TI cannot be made trivial for both phases without fine-tuning at order $V^2$.

In the 2D limit, a TPT requires a Zeeman energy comparable to the superconducting gap—an unphysical condition for vortices lacking magnetic impurities when $H \ll H_{c2}$. While a finite Zeeman field does couple the TPT to the effective chemical potential~\cite{fu2008superconducting,sau2010generic} ($E_F+\delta E_F$) and thereby to the CDW phase, this mechanism produces phenomena inconsistent with Ref.~\cite{liu2026intertwined} (as elaborated in the \ref{appd}). We therefore discard this model hereafter.

\section{Inversion symmetry breaking mechanism}
In the preceding analysis, the topological character stems from the TI band, while the superconducting order remains conventionally trivial. 
In contrast,
we propose a symmetry-based route by which a node-centered CDW can locally break inversion
symmetry, enable singlet–triplet mixing, and thereby favor an intrinsic topological vortex regime, where the nontrivial topology is inherent to the pairing symmetry itself.
Hereafter, we assume the band parameter modulation is insufficient to drive a TPT.
Crucially, unlike the mechanism discussed above, this transition is governed by the symmetry properties of the CDW rather than the fine-tuning of Hamiltonian parameters. Consequently, it yields a robust TPT that is symmetric for both sine and minus-sine CDW configurations.

\begin{figure}[t]
    \centering
    \includegraphics[width=0.45\textwidth]{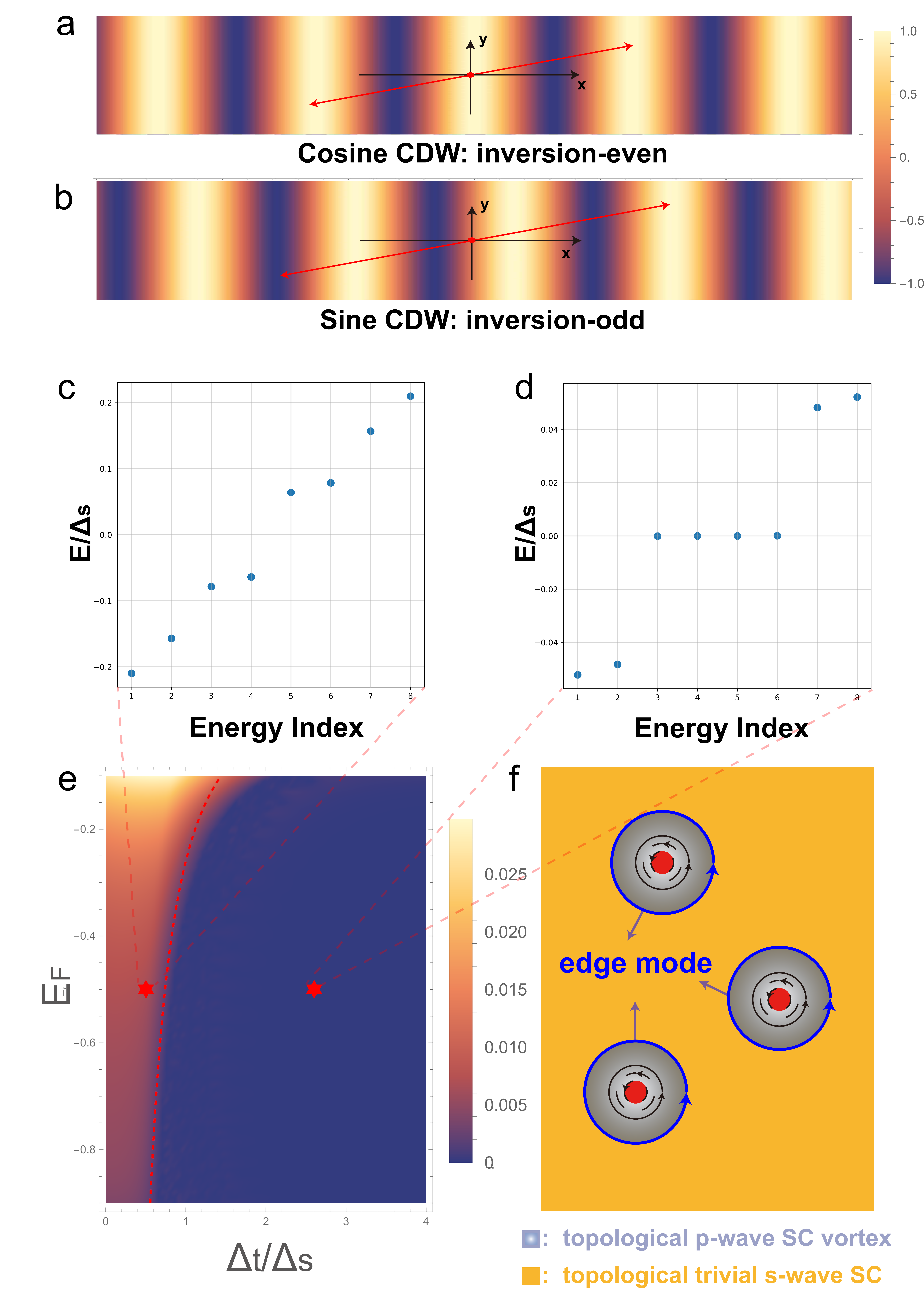}
    \caption{\textbf{(a)-(b)}: Schematic diagram of the preserving and breaking of inversion symmetry with cosine and sine form CDW. respectively. The red arrows indicate positions that are inversion partners of the vortex core (red points). \textbf{(c)-(d)}: The energy spectrum of a vortex with $\Delta_s=0.1$, $E_F=-0.5$. We choose the spin-triplet pairing order parameter as $\Delta_t=0.05$ in (c) and $\Delta_t=0.25$ in (d). Detailed discussion about the spectrum can be find in \ref{appf}. \textbf{(e)}: one half of the energy gap inside the vortex with different $E_F$ and $\Delta_t$. The red curve represents the condition $\Delta_t|\vec{d}|=\Delta_s$. As expected, on the right of the curve, energy gap approximately vanished, which indicated the appearance of MZM. We have label the parameter of (c) and (d) in the phase diagram. In this calculation, the system involves $60\times60$ lattices. The parameters are: $\Delta_s=0.1$, $v_F=0.1$, $\xi=2$, $Q=2\pi/3$, $m'=0$ and $\epsilon'=0.25$. The calculation is performed with the kwant package~\cite{groth2014kwant}. \textbf{(f)}: Schematic diagram of the topological $p$-wave SC vortices (in which the $p$-wave pairing component is larger than that of s-wave pairing) inside a 2D s-wave SC. Since the vortices (blue, encircled by the solid black lines) are topological nontrivial while the s-wave SC (yellow) is topologically trivial, there is expected to be chiral Majorana edge mode (represented by solid blue lines) at the boundaries between the two regions.}
    \label{triplet}
\end{figure}

Comparing sine- and cosine-form CDWs reveals a fundamental symmetry distinction: the cosine modulation preserves inversion symmetry, $\mathcal{I}H_{CDW}(\vec{r})\mathcal{I}^{-1}\equiv H_{CDW}(-\vec{r})=H_{CDW}(\vec{r})$, whereas the sine modulation breaks it (as shown in Fig.~\hyperref[triplet]{3(a)-(b)}). This inversion symmetry breaking (ISB) destroys parity conservation, permitting the admixture of spin-singlet and spin-triplet superconducting order parameters. The total order parameter is then $\Delta=\Delta_s+\Delta_t\vec{d}\cdot\vec{\sigma}$. In noncentrosymmetric systems, the triplet $\vec{d}$-vector aligns with the antisymmetric spin-orbit coupling (SOC) vector $\vec{L}$ in the SOC Hamiltonian $H_{SOC}=\alpha \vec{L}\cdot \vec{\sigma}$(see review in \ref{appe}), where SOC origins from an ISB potential $P(\vec{r})$: $\alpha\propto |\vec{\nabla}P(\vec{r})|$~\cite{frigeri2004superconductivity,sato2009topological,smidman2017superconductivity,zhang2025spin}.
In the present context, the effective ISB strength is strengthened by the potential gradient at the core, $VQ\cos\varphi$.
Moreover, the magnitude of this induced triplet component scales with the strength of SOC.
Physically, SOC breaks the degeneracy of spin, therefore introduce the difference in the Fermi-level density of states between the spin-split bands, $N_+-N_-\propto \alpha$. Furthermore, as the microscopic theory established~\cite{samokhin2008gap, bauer2012non}, the ratio of SC components satisfies:
\begin{equation}
r\equiv \Delta_t|\vec{d}|/\Delta_s\propto N_+-N_-\propto\alpha=\alpha_0+\beta VQ\cos\varphi,
\end{equation}
with $\alpha_0$ the bare SOC strength and $\beta$ a prefactor depending on the system details. Consequently, as the CDW evolves from a cosine to a sine profile, the ISB (and thus the $p$-wave component) is enhanced.

We first focus on the 3D model. To rule out the standard TI TPT mechanism in Eq.~\eqref{3Dmodel}, we assume a chemical potential large enough to render the unperturbed vortex topologically trivial. Driven by the Rashba SOC in $H_{TI}$, an inversion-symmetry-breaking CDW induces a $p$-wave superconducting state with the $d$-vector:
\begin{equation}\label{3Ddvector}
    \vec{d}_{3D}=(p_x,p_y,p_z).
\end{equation}
Viewing the vortex as a quasi-1D system with $p_z$ as a good quantum number, the system at each $p_z$ acts as a 2D helical $p$-wave superconductor with $\vec{d}=(p_x, p_y, 0)$ hosting its own CdGM states. However, the 3D solutions actually form a continuous spectrum due to the $p_z$-dependent dispersion~\cite{silaev2010topological,nishida2010color}. Moreover, the lack of a bulk insulating gap along the vortex line allows the expected surface MZMs to couple via these extended modes and split (see numerical result in \ref{appg} Information~\cite{supplement}). As a result, the system fails to produce a discrete topological CdGM spectrum. Thus, we discard this 3D model.

We then turn to the 2D limit, which is consistent with the ultrathin samples in Ref.~\cite{liu2026intertwined}.
Since the intrinsic Rashba SOC $v_F(p_x\sigma_x+p_y\sigma_y)$ dominates, the ISB potential will enforce a topological vector structure:
\begin{equation}\label{dvector}
    \vec{d}_{2D}=(p_x,p_y,0).
\end{equation}
It is well-established that spin-triplet pairing derived from Rashba SOC [Eq.~\eqref{dvector}] is topological nontrivial~\cite{sato2009topological,wang2017topological,zhang2025spin}. Specifically, theoretical analysis~\cite{sato2009topological} indicates that if the bulk gap remains open and the triplet amplitude dominates the singlet component on the Fermi surface ($\Delta_t|\vec{d}_{2D}|>\Delta_s$), the corresponding vortex-free system Hamiltonian acquires a nonzero $\mathbb{Z}_2$ invariant, characterizing a topological $p$-wave superconductor. Consequently, as the CDW configuration approaches the sine form, the enhanced ISB boosts the topological triplet component. Once the dominance condition is met, a TPT is triggered in the vicinity of $\varphi \approx 0$ or $\pi$. To test the topological nature of this mixed-parity state, we calculate the vortex energy spectrum for varying $s$-wave and $p$-wave admixtures, as shown in Fig.~\hyperref[triplet]{3(c)-(e)}. Consistent with expectations, increasing the triplet fraction drives the transition from a conventional vortex to a topological one hosting zero modes. This evolution is summarized in the phase diagram in Fig.~\hyperref[triplet]{3(e)}: when $\Delta_t|\vec{d}_{2D}|>\Delta_s$, the minigap vanishes, signaling the emergence of MZMs. We thus conclude that a sine-form CDW acts as a tunable symmetry-breaking field, enhancing the ISB sufficiently to induce a TPT within an initially conventional vortex. 

While singlet-triplet mixing in noncentrosymmetric systems is established~\cite{sato2009topological}, our model provides a distinct, local pathway to topological superconductivity. Unlike bulk spin-triplet materials~\cite{ran2019nearly,aoki2022unconventional,yang2021spin}, global $p$-wave pairing is unnecessary for vortex MZMs; the uniquely defined inversion center of the vortex core enables controlled local symmetry breaking. Because this noncentrosymmetry stems from an ISB potential, $p$-wave signatures can be manipulated without complex artificial tuning. Furthermore, the correlation between the CDW configuration and the topological state offers an experimentally accessible probe for intrinsic $p$-wave components.

Moreover, strong ISB inducing a large $p$-wave component is compatible with our weak-CDW assumption. In multi-band iron-based superconductors, a CDW may originate in the bulk bands, generating sufficient ISB to mix parity states and reconstruct the superconducting pairing. Providing the coupling between these bulk bands and the surface states, which governs the CdGM spectrum, is weak, the effective potential $H_{CDW}$ can remain perturbative.
In the meanwhile, if the bare SOC lies marginally below the topological phase transition, a perturbative CDW can effectively enhance the SOC, driving the vortex into the topological regime. With limited data, extracting the band parameters and CDW strength in Ref.~\cite{liu2026intertwined} is difficult; we defer quantitative values to future theory-experiment work.

\section{Experimental outlook}
To conclude, as summarized in Table~\ref{table1}, the 2D ISB mechanism is robust and consistent with symmetry. The corresponding predicts that vortex topology is controlled primarily by the local inversion breaking at the vortex core induced by a node-centered (sine-form) CDW. The failure of the corresponding 3D model highlights the crucial role of reduced dimensionality, suggesting the capacity to realize effective $p$-wave pairing and MZMs using artificial ISB potentials in 2D devices.
Since conventional bulk probes of spin-triplet pairing (e.g., Knight shift) are not vortex-resolved, identifying spatially local signatures is paramount.

Because topological superconductivity is confined to the vortex, it acts as an isolated topological island. Consequently, additional low-energy spectral weight near the vortex perimeter is expected~\cite{qi2010chiral}, as illustrated in Fig.~\hyperref[triplet]{3(f)}. High-resolution scanning tunneling microscope (STM)/scanning tunneling spectroscopy (STS) can test this: a zero-bias anomaly accompanied by a ring-like perimeter spectrum should emerge exclusively for node-centered CDW vortices.

A second discriminator is the Fermi level dependence. In our ISB framework, the Rashba-enhanced triplet component scales with Fermi momentum ($|\vec{d}| \sim k_F$); therefore, increasing $|E_F|$ expands the topological regime, as shown in Fig.~\hyperref[triplet]{3(e)}. Consequently, tuning the chemical potential via gating or doping should systematically strengthen these topological signatures.

Broadly, this mechanism provides a practical control knob: any engineered, vortex-centered inversion-breaking potential that preserves CdGM spectral coherence can toggle vortices in and out of the topological regime.

\begin{acknowledgments}
    \textbf{Acknowledgments}: The authors thank Can-Li Song for the fruitful discussion and for providing experimental data. Z.Z. and D.E.L. are supported by the Quantum Science and Technology-National Science and Technology Major Project (Grant No. 2021ZD0302400). Z.W. is supported by the U.S. Department of Energy, Basic Energy Sciences Grant DE-FG02-99ER45747.
\end{acknowledgments}

\bibliography{ref}

@article{kawakami2015evolution,
  title={Evolution of density of states and a spin-resolved checkerboard-type pattern associated with the Majorana bound state},
  author={Kawakami, Takuto and Hu, Xiao},
  journal={Physical Review Letters},
  volume={115},
  number={17},
  pages={177001},
  year={2015},
  publisher={APS},
  url={https://link.aps.org/doi/10.1103/PhysRevLett.115.177001}
}

@article{hosur2011majorana,
  title={Majorana modes at the ends of superconductor vortices in doped topological insulators},
  author={Hosur, Pavan and Ghaemi, Pouyan and Mong, Roger SK and Vishwanath, Ashvin},
  journal={Physical review letters},
  volume={107},
  number={9},
  pages={097001},
  year={2011},
  publisher={APS},
  url={https://journals.aps.org/prl/abstract/10.1103/PhysRevLett.107.097001}
}

@article{deng2020bound,
  title={Bound fermion states in pinned vortices in the surface states of a superconducting topological insulator},
  author={Deng, Haoyun and Bonesteel, Nicholas and Schlottmann, Pedro},
  journal={Journal of Physics: Condensed Matter},
  volume={33},
  number={3},
  pages={035604},
  year={2020},
  publisher={IOP Publishing},
  url={https://iopscience.iop.org/article/10.1088/1361-648X/abba89/meta}
}

@article{banerjee2025charge,
  title={Charge density wave solutions of the Hubbard model in the composite operator formalism},
  author={Banerjee, Anurag and Pangburn, Emile and Mahato, Chiranjit and Ghosal, Amit and P{\'e}pin, Catherine},
  journal={Physical Review B},
  volume={111},
  number={16},
  pages={165123},
  year={2025},
  publisher={APS},
  url={https://journals.aps.org/prb/abstract/10.1103/PhysRevB.111.165123}
}

@article{feng2021chiral,
  title={Chiral flux phase in the Kagome superconductor AV3Sb5},
  author={Feng, Xilin and Jiang, Kun and Wang, Ziqiang and Hu, Jiangping},
  journal={Science bulletin},
  volume={66},
  number={14},
  pages={1384--1388},
  year={2021},
  publisher={Elsevier},
  url={https://www.sciencedirect.com/science/article/pii/S2095927321003224}
}

@article{zeng2023chiral,
  title={Chiral-flux-phase-based topological superconductivity in kagome systems with mixed edge chiralities},
  author={Zeng, Junjie and Li, Qingming and Yang, Xun and Xu, Dong-Hui and Wang, Rui},
  journal={Physical Review Letters},
  volume={131},
  number={8},
  pages={086601},
  year={2023},
  publisher={APS},
  url={https://journals.aps.org/prl/abstract/10.1103/PhysRevLett.131.086601}
}

@article{lin2024impact,
  title={Impact of charge density waves on superconductivity and topological properties in kagome superconductors},
  author={Lin, Xin and Huang, Junkang and Zhou, Tao},
  journal={Physical Review B},
  volume={110},
  number={13},
  pages={134502},
  year={2024},
  publisher={APS},
  url={https://journals.aps.org/prb/abstract/10.1103/PhysRevB.110.134502}
}

@article{frigeri2004superconductivity,
  title={Superconductivity without Inversion Symmetry: MnSi versus CePt3Si},
  author={Frigeri, PA and Agterberg, DF and Koga, A and Sigrist, M},
  journal={Physical review letters},
  volume={92},
  number={9},
  pages={097001},
  year={2004},
  publisher={APS},
  url={https://journals.aps.org/prl/abstract/10.1103/PhysRevLett.92.097001}
}

@article{smidman2017superconductivity,
  title={Superconductivity and spin--orbit coupling in non-centrosymmetric materials: a review},
  author={Smidman, M and Salamon, MB and Yuan, HQ and Agterberg, DF},
  journal={Reports on Progress in Physics},
  volume={80},
  number={3},
  pages={036501},
  year={2017},
  publisher={IOP Publishing},
  url={https://iopscience.iop.org/article/10.1088/1361-6633/80/3/036501/meta}
}

@incollection{ginzburg2009theory,
  title={On the theory of superconductivity},
  author={Ginzburg, Vitaly L and Landau, Lev D},
  booktitle={On superconductivity and superfluidity: a scientific autobiography},
  pages={113--137},
  year={2009},
  publisher={Springer},
  url={https://link.springer.com/chapter/10.1007/978-3-540-68008-6_4}
}

@article{abrikosov1957magnetic,
  title={On the magnetic properties of superconductors of the second group},
  author={Abrikosov, Alexei A},
  journal={Soviet Physics-JETP},
  volume={5},
  pages={1174--1182},
  year={1957},
  url={https://elibrary.ru/item.asp?id=21757785}
}

@article{wu2013emergence,
  title={Emergence of charge order from the vortex state of a high-temperature superconductor},
  author={Wu, Tao and Mayaffre, Hadrien and Kr{\"a}mer, Steffen and Horvati{\'c}, Mladen and Berthier, Claude and Kuhns, Philip L and Reyes, Arneil P and Liang, Ruixing and Hardy, WN and Bonn, DA and Julien, Marc-Henri},
  journal={Nature communications},
  volume={4},
  number={1},
  pages={2113},
  year={2013},
  publisher={Nature Publishing Group UK London},
  url={https://www.nature.com/articles/ncomms3113}
}

@article{lake2002antiferromagnetic,
  title={Antiferromagnetic order induced by an applied magnetic field in a high-temperature superconductor},
  author={Lake, B and R{\o}nnow, HM and Christensen, NB and Aeppli, G and Lefmann, K and McMorrow, DF and Vorderwisch, P and Smeibidl, P and Mangkorntong, N and Sasagawa, T and Nohara, M and Takagi, H and Mason, T E},
  journal={Nature},
  volume={415},
  number={6869},
  pages={299--302},
  year={2002},
  publisher={Nature Publishing Group UK London},
  url={https://www.nature.com/articles/415299a}
}

@article{hoffman2002four,
  title={A four unit cell periodic pattern of quasi-particle states surrounding vortex cores in Bi2Sr2CaCu2O8+ $\delta$},
  author={Hoffman, Jennifer E and Hudson, Eric W and Lang, KM and Madhavan, Vidya and Eisaki, Hiroshi and Uchida, Shin’ichi and Davis, James C},
  journal={Science},
  volume={295},
  number={5554},
  pages={466--469},
  year={2002},
  publisher={American Association for the Advancement of Science},
  url={https://www.science.org/doi/abs/10.1126/science.1066974}
}

@article{katano2000enhancement,
  title={Enhancement of static antiferromagnetic correlations by magnetic field in a superconductor La 2- x Sr x CuO 4 with x= 0.12},
  author={Katano, Susumu and Sato, Masugu and Yamada, Kazuyoshi and Suzuki, Takao and Fukase, Tetsuo},
  journal={Physical Review B},
  volume={62},
  number={22},
  pages={R14677},
  year={2000},
  publisher={APS},
  url={https://journals.aps.org/prb/abstract/10.1103/PhysRevB.62.R14677}
}

@article{chang2008tuning,
  title={Tuning competing orders in La 2- x Sr x CuO 4 cuprate superconductors by the application of an external magnetic field},
  author={Chang, J. and Niedermayer, Ch. and Gilardi, R. and Christensen, N. B. and R\o{}nnow, H. M. and McMorrow, D. F. and Ay, M. and Stahn, J. and Sobolev, O. and Hiess, A. and Pailhes, S. and Baines, C. and Momono, N. and Oda, M. and Ido, M. and Mesot, J.},
  journal={Physical Review B—Condensed Matter and Materials Physics},
  volume={78},
  number={10},
  pages={104525},
  year={2008},
  publisher={APS},
  url={https://journals.aps.org/prb/abstract/10.1103/PhysRevB.78.104525}
}

@article{edkins2019magnetic,
  title={Magnetic field--induced pair density wave state in the cuprate vortex halo},
  author={Edkins, Stephen D and Kostin, Andrey and Fujita, Kazuhiro and Mackenzie, Andrew P and Eisaki, Hiroshi and Uchida, S and Sachdev, Subir and Lawler, Michael J and Kim, E-A and S{\'e}amus Davis, JC and Hamidian, M H},
  journal={Science},
  volume={364},
  number={6444},
  pages={976--980},
  year={2019},
  publisher={American Association for the Advancement of Science},
  url={https://www.science.org/doi/abs/10.1126/science.aat1773}
}

@article{caroli1964bound,
  title={Bound fermion states on a vortex line in a type II superconductor},
  author={Caroli, C and De Gennes, PG and Matricon, J},
  journal={Physics Letters},
  volume={9},
  number={4},
  pages={307--309},
  year={1964},
  publisher={Elsevier},
  url={https://www.sciencedirect.com/science/article/pii/0031916364903750}
}

@article{berthod2017observation,
  title={Observation of Caroli--de Gennes--Matricon vortex states in YBa 2 Cu 3 O 7- $\delta$},
  author={Berthod, Christophe and Maggio-Aprile, Ivan and Bru{\'e}r, Jens and Erb, Andreas and Renner, Christoph},
  journal={Physical review letters},
  volume={119},
  number={23},
  pages={237001},
  year={2017},
  publisher={APS},
  url={https://journals.aps.org/prl/abstract/10.1103/PhysRevLett.119.237001}
}

@article{chen2018discrete,
  title={Discrete energy levels of Caroli-de Gennes-Matricon states in quantum limit in FeTe0. 55Se0. 45},
  author={Chen, Mingyang and Chen, Xiaoyu and Yang, Huan and Du, Zengyi and Zhu, Xiyu and Wang, Enyu and Wen, Hai-Hu},
  journal={Nature communications},
  volume={9},
  number={1},
  pages={970},
  year={2018},
  publisher={Nature Publishing Group UK London},
  url={https://www.nature.com/articles/s41467-018-03404-8}
}

@article{liu2018robust,
  title={Robust and clean Majorana zero mode in the vortex core of high-temperature superconductor (Li 0.84 Fe 0.16) OHFeSe},
  author={Liu, Qin and Chen, Chen and Zhang, Tong and Peng, Rui and Yan, Ya-Jun and Wen, Chen-Hao-Ping and Lou, Xia and Huang, Yu-Long and Tian, Jin-Peng and Dong, Xiao-Li and Wang, Guang-Wei and Bao, Wei-Cheng and Wang, Qiang-Hua and Yin, Zhi-Ping and Zhao, Zhong-Xian and Feng, Dong-Lai},
  journal={Physical Review X},
  volume={8},
  number={4},
  pages={041056},
  year={2018},
  publisher={APS},
  url={https://journals.aps.org/prx/abstract/10.1103/PhysRevX.8.041056}
}

@article{nayak2008non,
  title={Non-Abelian anyons and topological quantum computation},
  author={Nayak, Chetan and Simon, Steven H and Stern, Ady and Freedman, Michael and Das Sarma, Sankar},
  journal={Reviews of Modern Physics},
  volume={80},
  number={3},
  pages={1083--1159},
  year={2008},
  publisher={APS},
  url={https://journals.aps.org/rmp/abstract/10.1103/RevModPhys.80.1083}
}

@article{fu2008superconducting,
  title={Superconducting proximity effect and majorana fermions at the surface of a topological insulator},
  author={Fu, Liang and Kane, Charles L},
  journal={Physical review letters},
  volume={100},
  number={9},
  pages={096407},
  year={2008},
  publisher={APS},
  url={https://journals.aps.org/prl/abstract/10.1103/PhysRevLett.100.096407}
}

@article{zhang2018observation,
  title={Observation of topological superconductivity on the surface of an iron-based superconductor},
  author={Zhang, Peng and Yaji, Koichiro and Hashimoto, Takahiro and Ota, Yuichi and Kondo, Takeshi and Okazaki, Kozo and Wang, Zhijun and Wen, Jinsheng and Gu, Genda D and Ding, Hong and Shin, Shik},
  journal={Science},
  volume={360},
  number={6385},
  pages={182--186},
  year={2018},
  publisher={American Association for the Advancement of Science},
  url={https://www.science.org/doi/abs/10.1126/science.aan4596}
}

@article{zhang2021observation,
  title={Observation of distinct spatial distributions of the zero and nonzero energy vortex modes in (Li 0.84 Fe 0.16) OHFeSe},
  author={Zhang, Tianzhen and Bao, Weicheng and Chen, Chen and Li, Dong and Lu, Zouyuwei and Hu, Yining and Yang, Wentao and Zhao, Dongming and Yan, Yajun and Dong, Xiaoli and Wang, Qiang-Hua and Zhang, Tong and Feng, Donglai},
  journal={Physical Review Letters},
  volume={126},
  number={12},
  pages={127001},
  year={2021},
  publisher={APS},
  url={https://journals.aps.org/prl/abstract/10.1103/PhysRevLett.126.127001}
}

@article{kong2021majorana,
  title={Majorana zero modes in impurity-assisted vortex of LiFeAs superconductor},
  author={Kong, Lingyuan and Cao, Lu and Zhu, Shiyu and Papaj, Micha{\l} and Dai, Guangyang and Li, Geng and Fan, Peng and Liu, Wenyao and Yang, Fazhi and Wang, Xiancheng and Du, Shixuan and Jin, Changqing and Fu, Liang and Gao, Hong-Jun and Ding, Hong},
  journal={Nature Communications},
  volume={12},
  number={1},
  pages={4146},
  year={2021},
  publisher={Nature Publishing Group UK London},
  url={https://www.nature.com/articles/s41467-021-24372-6}
}

@article{li2022ordered,
  title={Ordered and tunable Majorana-zero-mode lattice in naturally strained LiFeAs},
  author={Li, Meng and Li, Geng and Cao, Lu and Zhou, Xingtai and Wang, Xiancheng and Jin, Changqing and Chiu, Ching-Kai and Pennycook, Stephen J and Wang, Ziqiang and Gao, Hong-Jun},
  journal={Nature},
  volume={606},
  number={7916},
  pages={890--895},
  year={2022},
  publisher={Nature Publishing Group UK London},
  url={https://www.nature.com/articles/s41586-022-04744-8}
}

@article{machida2019zero,
  title={Zero-energy vortex bound state in the superconducting topological surface state of Fe (Se, Te)},
  author={Machida, T and Sun, Y and Pyon, S and Takeda, S and Kohsaka, Y and Hanaguri, T and Sasagawa, T and Tamegai, T},
  journal={Nature materials},
  volume={18},
  number={8},
  pages={811--815},
  year={2019},
  publisher={Nature Publishing Group UK London},
  url={https://www.nature.com/articles/s41563-019-0397-1}
}

@article{sato2009topological,
  title={Topological phases of noncentrosymmetric superconductors: Edge states, Majorana fermions, and non-Abelian statistics},
  author={Sato, Masatoshi and Fujimoto, Satoshi},
  journal={Physical Review B—Condensed Matter and Materials Physics},
  volume={79},
  number={9},
  pages={094504},
  year={2009},
  publisher={APS},
  url=https://journals.aps.org/prb/abstract/10.1103/PhysRevB.79.094504}

@article{chiu2020scalable,
  title={Scalable Majorana vortex modes in iron-based superconductors},
  author={Chiu, Ching-Kai and Machida, T and Huang, Yingyi and Hanaguri, T and Zhang, Fu-Chun},
  journal={Science Advances},
  volume={6},
  number={9},
  pages={eaay0443},
  year={2020},
  publisher={American Association for the Advancement of Science},
  url={https://www.science.org/doi/abs/10.1126/sciadv.aay0443}
}

@article{abrikosov2004nobel,
  title={Nobel Lecture: Type-II superconductors and the vortex lattice},
  author={Abrikosov, Aleksej A},
  journal={Reviews of modern physics},
  volume={76},
  number={3},
  pages={975--979},
  year={2004},
  publisher={APS},
  url={https://journals.aps.org/rmp/abstract/10.1103/RevModPhys.76.975}
}

@article{hu2016theory,
  title={Theory of spin-selective Andreev reflection in the vortex core of a topological superconductor},
  author={Hu, Lun-Hui and Li, Chuang and Xu, Dong-Hui and Zhou, Yi and Zhang, Fu-Chun},
  journal={Physical Review B},
  volume={94},
  number={22},
  pages={224501},
  year={2016},
  publisher={APS},
  url={https://link.aps.org/doi/10.1103/PhysRevB.94.224501}
}

@article{rakhmanov2011majorana,
  title={Majorana fermions in pinned vortices},
  author={Rakhmanov, AL and Rozhkov, AV and Nori, Franco},
  journal={Physical Review B—Condensed Matter and Materials Physics},
  volume={84},
  number={7},
  pages={075141},
  year={2011},
  publisher={APS},
  url={https://journals.aps.org/prb/abstract/10.1103/PhysRevB.84.075141}
}

@article{banerjee2022charge,
  title={Charge, bond, and pair density wave orders in a strongly correlated system},
  author={Banerjee, Anurag and P{\'e}pin, Catherine and Ghosal, Amit},
  journal={Physical Review B},
  volume={105},
  number={13},
  pages={134505},
  year={2022},
  publisher={APS},
  url={https://journals.aps.org/prb/abstract/10.1103/PhysRevB.105.134505}
}

@article{zhang2025spin,
  title={Spin-triplet pair density wave superconductors},
  author={Zhang, Yi and Wang, Ziqiang},
  journal={Communications Physics},
  volume={8},
  number={1},
  pages={337},
  year={2025},
  publisher={Nature Publishing Group UK London},
  url={https://www.nature.com/articles/s42005-025-02256-1}
}

@article{groth2014kwant,
  title={Kwant: a software package for quantum transport},
  author={Groth, Christoph W and Wimmer, Michael and Akhmerov, Anton R and Waintal, Xavier},
  journal={New Journal of Physics},
  volume={16},
  number={6},
  pages={063065},
  year={2014},
  publisher={IOP Publishing},
  url={https://iopscience.iop.org/article/10.1088/1367-2630/16/6/063065/meta}
}

@article{wang2017topological,
  title={Topological phase transitions in multicomponent superconductors},
  author={Wang, Yuxuan and Fu, Liang},
  journal={Physical review letters},
  volume={119},
  number={18},
  pages={187003},
  year={2017},
  publisher={APS},
  url={https://journals.aps.org/prl/abstract/10.1103/PhysRevLett.119.187003}
}

@article{read2000paired,
  title={Paired states of fermions in two dimensions with breaking of parity and time-reversal symmetries and the fractional quantum Hall effect},
  author={Read, Nicholas and Green, Dmitry},
  journal={Physical Review B},
  volume={61},
  number={15},
  pages={10267},
  year={2000},
  publisher={APS},
  url={https://journals.aps.org/prb/abstract/10.1103/PhysRevB.61.10267}
}

@article{kitaev2001unpaired,
  title={Unpaired majorana fermions in quantumwires},
  author={Kitaev, A Yu},
  journal={Physics-uspekhi},
  volume={44},
  number={10S},
  pages={131},
  year={2001},
  publisher={IOP Publishing},
  url={https://iopscience.iop.org/article/10.1070/1063-7869/44/10S/S29/meta}
}

@article{ran2019nearly,
  title={Nearly ferromagnetic spin-triplet superconductivity},
  author={Ran, Sheng and Eckberg, Chris and Ding, Qing-Ping and Furukawa, Yuji and Metz, Tristin and Saha, Shanta R and Liu, I-Lin and Zic, Mark and Kim, Hyunsoo and Paglione, Johnpierre and Butch, Nicholas P.},
  journal={Science},
  volume={365},
  number={6454},
  pages={684--687},
  year={2019},
  publisher={American Association for the Advancement of Science},
  url={https://www.science.org/doi/abs/10.1126/science.aav8645}
}

@article{yang2021spin,
  title={Spin-triplet superconductivity in K2Cr3As3},
  author={Yang, Jie and Luo, Jun and Yi, Changjiang and Shi, Youguo and Zhou, Yi and Zheng, Guo-qing},
  journal={Science advances},
  volume={7},
  number={52},
  pages={eabl4432},
  year={2021},
  publisher={American Association for the Advancement of Science},
  url={https://www.science.org/doi/abs/10.1126/sciadv.abl4432}
}

@article{aoki2022unconventional,
  title={Unconventional superconductivity in UTe2},
  author={Aoki, Dai and Brison, Jean-Pascal and Flouquet, Jacques and Ishida, Kenji and Knebel, Georg and Tokunaga, Y and Yanase, Youichi},
  journal={Journal of Physics: Condensed Matter},
  volume={34},
  number={24},
  pages={243002},
  year={2022},
  publisher={IOP Publishing},
  url={https://iopscience.iop.org/article/10.1088/1361-648X/ac5863/meta}
}

@article{liu2026intertwined,
  title={Intertwined Charge Stripes and Majorana Zero Modes in An Iron-Based Superconductor},
  author={Liu, Yu and Wei, Li-Xuan and Cheng, Qiang-Jun and Zhu, Zhenhua and Shi, Xin-Yu and Lou, Cong-Cong and Wang, Yong-Wei and Deng, Ze-Xian and Ren, Ming-Qiang and Liu, Dong E and Wang, Ziqiang and Ma, Xu-Cun and Jia, Jin-Feng and Xue, Qi-Kun and Song, Can-Li},
  journal={arXiv preprint arXiv:2601.15873},
  year={2026},
  url={https://arxiv.org/abs/2601.15873}
}

@article{wang2018evidence,
  title={Evidence for Majorana bound states in an iron-based superconductor},
  author={Wang, Dongfei and Kong, Lingyuan and Fan, Peng and Chen, Hui and Zhu, Shiyu and Liu, Wenyao and Cao, Lu and Sun, Yujie and Du, Shixuan and Schneeloch, John and Zhong, Ruidan and Gu, Genda and Fu, Liang and Ding, Hong and Gao, Hong-Jun},
  journal={Science},
  volume={362},
  number={6412},
  pages={333--335},
  year={2018},
  publisher={American Association for the Advancement of Science},
  url={https://www.science.org/doi/abs/10.1126/science.aao1797}
}

@inbook{bauer2012non,
  title={Basic Theory of Superconductivity in Metals Without Inversion Center},
  booktitle={Non-centrosymmetric superconductors: introduction and overview},
  pages={129–154},
  author={Bauer, Ernst and Sigrist, Manfred},
  volume={847},
  year={2012},
  publisher={Springer Science \& Business Media},
  url={https://books.google.com/books?hl=zh-CN&lr=&id=nDZ4lKD00t8C&oi=fnd&pg=PR5&dq=Non-Centrosymmetric+Superconductors&ots=G6iw4KFs8u&sig=sJ73E2tptEyEzUuERNWsmW-hdxc}}

@article{qi2010chiral,
  title={Chiral topological superconductor from the quantum Hall state},
  author={Qi, Xiao-Liang and Hughes, Taylor L and Zhang, Shou-Cheng},
  journal={Physical Review B—Condensed Matter and Materials Physics},
  volume={82},
  number={18},
  pages={184516},
  year={2010},
  publisher={APS},
  url={https://journals.aps.org/prb/abstract/10.1103/PhysRevB.82.184516}
}

@article{samokhin2008gap,
  title={Gap structure in noncentrosymmetric superconductors},
  author={Samokhin, KV and Mineev, VP},
  journal={Physical Review B—Condensed Matter and Materials Physics},
  volume={77},
  number={10},
  pages={104520},
  year={2008},
  publisher={APS},
  url={https://journals.aps.org/prb/abstract/10.1103/PhysRevB.77.104520}
}

@article{sau2010generic,
  title={Generic new platform for topological quantum computation using semiconductor heterostructures},
  author={Sau, Jay D and Lutchyn, Roman M and Tewari, Sumanta and Das Sarma, Sankar},
  journal={Physical review letters},
  volume={104},
  number={4},
  pages={040502},
  year={2010},
  publisher={APS},
  url={https://journals.aps.org/prl/abstract/10.1103/PhysRevLett.104.040502}
}

@article{silaev2010topological,
  title={Topological Superfluid 3He--B: Fermion Zero Modes on Interfaces and in the Vortex Core},
  author={Silaev, MA and Volovik, Grigory E},
  journal={Journal of Low Temperature Physics},
  volume={161},
  number={5},
  pages={460--473},
  year={2010},
  publisher={Springer},
  url={https://link.springer.com/article/10.1007/s10909-010-0226-z}
}

@article{nishida2010color,
  title={Is a color superconductor topological?},
  author={Nishida, Yusuke},
  journal={Physical Review D—Particles, Fields, Gravitation, and Cosmology},
  volume={81},
  number={7},
  pages={074004},
  year={2010},
  publisher={APS},
  url={https://journals.aps.org/prd/abstract/10.1103/PhysRevD.81.074004}
}

@misc{supplement,
title = {See Supplemental materials for details of numerical results.},
note = {}
}

\appendix

\section{Details of CDW in the 3D model}\label{appa}
Here we detail the implementation of CDW effects within the 3D Hamiltonian. As outlined in the main text, the onsite component (oCDW) modulates the local chemical potential, $E_F'(\vec{r})=E_{F}+V_o\sin(Qx+\varphi)$, while the hopping component (hCDW) renormalizes the nearest-neighbor hopping amplitude, $\epsilon(\vec{r})=\epsilon+V_h\sin(Qx+\varphi)$. Incorporating these into the continuum limit, the full CDW perturbation in Eq.~\eqref{3Dmodel} takes the form:
\begin{equation}
H_{CDW}^{(3D)}=[(6-p^2)V_h\tau_z - V_o\tau_0]f(\vec{r})\sin(Qx+\varphi).
\end{equation}
To clarify the TPT mechanism, we treat $H_{CDW}$ perturbatively. When a CDW node coincides with the vortex core (sine configuration), the potential vanishes locally ($\sin(0)=0$). Furthermore, symmetry dictates that the first-order energy correction,
\begin{equation}
\delta E_\mu^{(1)} = \int d^3r\, \Psi_\mu^\dagger H_{o(h)CDW} \Psi_\mu,
\end{equation}
must vanish, as the integrand combines an inversion-even density $|\Psi_\mu|^2$ with an inversion-odd potential. Conversely, when a CDW antinode is centered at the core (cosine configuration), the potential is locally non-vanishing and inversion-symmetric, allowing a finite first-order correction that shifts the effective band parameters

\section{Phases of oCDW and hCDW}\label{appb}
In the preceding analysis of direct parameter modulation, we enforced phase locking between the onsite and hopping CDW components ($\varphi_o = \varphi_h$). This constraint is not arbitrary; it is the necessary condition to reconcile the model with the experimental observation of trivial vortices in both cosine and minus-cosine configurations. Consider, for instance, the antiphase case ($\varphi_o - \varphi_h = \pi$): a minus-cosine configuration would simultaneously suppress the local chemical potential and elevate the topological threshold $E_C$. These effects would constructively reinforce the topological stability condition ($E_F < E_C$), precluding the experimentally observed trivial state. A systematic survey of the parameter space confirms that only the in-phase configuration allows for the suppression of the topological phase in both modulation limits.

\section{Derivation of Eq.~\eqref{TPTcondition}}\label{appc}
Here we explicitly derive the constraints in Eq.~\eqref{TPTcondition} by accounting for corrections up to second order in the CDW strength. Let the modulations of the chemical potential and effective mass parameter induced by the cosine form CDW be denoted as
\begin{align}
\delta E_F &= \delta_1^{(1)} + \delta_1^{(2)},\\
\delta \epsilon &= \delta_2^{(1)} + \delta_2^{(2)},
\end{align}
where the superscript $(n)$ indicates a correction of order $V^n$. Consequently, for the minus-cosine configuration, the first-order corrections invert sign while the second-order terms remain invariant:
\begin{align}
\delta' E_F &= -\delta_1^{(1)} + \delta_1^{(2)},\\
\delta' \epsilon& = -\delta_2^{(1)} + \delta_2^{(2)}.
\end{align}
To simultaneously drive the vortex into the trivial regime for both CDW phases, the system must satisfy the following inequalities:
\begin{equation}
\begin{split}
E_F &< v_F\sqrt{m/\epsilon}, \\
E_F + \delta E_F &> v_F\sqrt{(m+6\delta\epsilon)/(\epsilon+\delta\epsilon)}, \\
E_F + \delta'E_F &> v_F\sqrt{(m+6\delta'\epsilon)/(\epsilon+\delta'\epsilon)}.
\end{split}
\end{equation}
It is straightforward to show that at first order, this system is inconsistent (see Eq.~\eqref{TPTcondition}. A solution emerges only upon including second-order corrections, which impose a strict lower bound on the chemical potential:
\begin{equation}\label{TPTcondition_derived}
\begin{split}
E_F &< v_F\sqrt{\frac{m}{\epsilon}}, \\
E_F > v_F\sqrt{\frac{m}{\epsilon}} - \delta_1^{(2)} &- A\delta_2^{(2)} + B[\delta^{(1)}]^2 + \mathcal{O}(V^3),
\end{split}
\end{equation}
where we have defined the constants $A \equiv (m-6\epsilon)/\sqrt{4m\epsilon^3}$ and $B \equiv 3(m-6\epsilon)(m+2\epsilon)/\sqrt{64m^3\epsilon^5}$. This demonstrates that the direct modulation mechanism can account for the experimental findings in Ref.~\cite{liu2026intertwined} only if the chemical potential is globally fine-tuned with a precision comparable to the second-order CDW energy corrections, and therefore is non-robust.

\section{Zeeman effect in the 2D model}\label{appd}
Incorporating a Zeeman term in Eq.~\eqref{2Dmodel}, the BdG Hamiltonian takes the form:
\begin{equation}\label{2Dmodel_end matter}
\begin{split}
H_{2D}^Z=\frac{1}{2}\int dr^2\Psi_{\vec{r}}^\dag \begin{pmatrix}
H_2+M_B\sigma_z & \hat{\Delta}(\vec{r})\\
\hat{\Delta}^\dag(\vec{r}) & -H_2+M_B\sigma_z
\end{pmatrix}\Psi_{\vec{r}}
\end{split},
\end{equation}
where $M_B$ denotes the Zeeman energy induced by a magnetic field applied parallel to the $z$-axis, and
\begin{equation}
H_2=H_{bulk}+v_F\vec{\sigma}\cdot\vec{p}-E_F.
\end{equation}
Here, we have omitted the CDW potential to focus on the topological properties of the system. Since the presence of the bulk term $H_{bulk}$ significantly impacts the topological classification, we analyze this model across two regimes:

If $H_{bulk}=0$, topological Dirac cone will protect the MZM inside the vortex when $M_B=0$. Introducing a Zeeman field opens a gap at the Dirac point, modulating the topological properties and governing the stability of the MZM. Specifically, a Fermi surface falling within the insulating gap will cause the MZM to vanish, signaling a TPT~\cite{fu2008superconducting}. However, a difficulty emerges when attempting to couple this TPT to the CDW modulation: for a Fermi surface (with positive energy) outside the insulating gap, while oCDW in the minus-cosine region locally lowers the effective chemical potential to induce a trivial vortex, the cosine region pushes the vortex deeper into the topological regime. Furthermore, relating the TPT to the ISB mechanism requires assuming that the chemical potential $E_F$ lies within the magnetic gap. This assumption results in a large CdGM level spacing, $\Delta_0^2/E_F$, a feature that is absent in the experimental data of Ref.~\cite{liu2026intertwined}. Consequently, this model cannot reproduce the phenomenology observed in Ref.~\cite{liu2026intertwined}.

Turning to the case where $H_{bulk}=-\epsilon'p^2$, the gap due to Zeeman field is opened in the Rashba-like band. the Zeeman field opens a gap in the Rashba-like band. Previous studies have established that the MZM only survives when $|E_F^{eff}|<\sqrt{M_B^2-\Delta_0^2}$~\cite{sau2010generic}. Assuming the bare chemical potential lies within this topological regime, oCDW in the cosine and minus-cosine regions could both theoretically drive the effective chemical potential into the trivial phase. However, this scenario presents two major inconsistencies with the observations in Ref.~\cite{liu2026intertwined}: 1) Since the effective chemical potential is modulated by oCDW, maintaining the perturbative nature of the oCDW requires the effective chemical potential to be significantly smaller than the CdGM level spacing. Consequently, the corresponding minigap $\Delta_0^2/E_F^{eff}$ would exceed the bulk superconducting gap, contradicting the experimental data. 2) In this paradigm, the effective chemical potential of topological vortices must be larger than that of trivial vortices. This implies that the CdGM level spacing for the two vortex types should be distinctly different, which also conflicts with the experimental findings.

Moreover, the aforementioned models require a Zeeman energy comparable to the superconducting gap $\Delta_0$. Under realistic experimental conditions, however, the Zeeman energy is much smaller than $\Delta_0$, rendering its effect on the vortex core topology negligible. Consequently, we neglect the Zeeman field in this work.

\section{Brief review in ISB induced p-wave SC theory}\label{appe}
Following Ref.~\cite{frigeri2004superconductivity}, we briefly review the origin of the spin-triplet pairing component in the presence of inversion symmetry breaking. In a system lacking inversion symmetry, the microscopic Hamiltonian,
\begin{equation}
\begin{split}
H=&\sum_{\vec{p},s} \epsilon_{\vec{p}}c_{\vec{p}s}^\dag
c_{\vec{p}s}+\frac{1}{2}\sum_{\vec{p},\vec{p}'}\sum_{s,s'}V_{\vec{p},\vec{p}'}c_{\vec{p}s}^\dag c_{-\vec{p}s'}^\dag c_{-\vec{p}'s}c_{\vec{p}'s}\\
&+\alpha\sum_{\vec{p}ss'}(\vec{L}_{\vec{p}}\cdot \vec{\sigma})_{ss'}c_{\vec{p}s}^\dag c_{\vec{p}s},
\end{split}
\end{equation}
allows for the admixture of spin-singlet and spin-triplet SC pairings. Here, the final term represents the SOC. By evaluating the transition temperature $T_c$ for the spin-triplet component, one obtains:
\begin{equation}
\ln(\frac{T_C}{T_0})=2\langle \{|\vec{d}|^2-|\vec{L}\cdot \vec{d}|^2\}f_{\vec{L}}\rangle+\mathcal{O}(\frac{\alpha^2}{\epsilon_F^2}),
\end{equation}
where $\vec{d}$ is the corresponding $d$-vector of the spin-triplet component, $f_{\vec{L}}$ is a function dependent on the SOC, $\mathcal{O}(x^n)$ denotes the correction terms up to the $n$-th order of the perturbation, and $\epsilon_F$ is the Fermi energy. The parameter $T_0$ is a constant determined by the bare Hamiltonian parameters. To energetically stabilize the induced spin-triplet state, $T_c$ must be maximized, leading to the condition $\vec{d}\parallel \vec{L}$.

\section{Notes on Fig.~\hyperref[triplet]{3(e)} and Fig.~\hyperref[triplet]{3(d)}}\label{appf}
Fig.~\hyperref[triplet]{3(e)} demonstrates robust agreement between the numerically extracted vortex gap and the analytical TPT condition. Deviations observed near $E_F=0$ are intrinsic to the breakdown of the quasi-classical approximation, as the standard CdGM description is strictly valid only in the limit $|E_F| \gg E_\mu$~\cite{deng2020bound}. Additionally, the spectral fluctuations near the phase boundary arise from finite-size effects: as the BdG gap closes at the transition, the coherence length of the bound state diverges, causing the wavefunction to overlap with the system boundaries. These artifacts notwithstanding, the global topology of the phase diagram confirms the theoretical predictions derived in the main text.

\begin{figure}[h]
    \centering
    \includegraphics[width=0.3\linewidth]{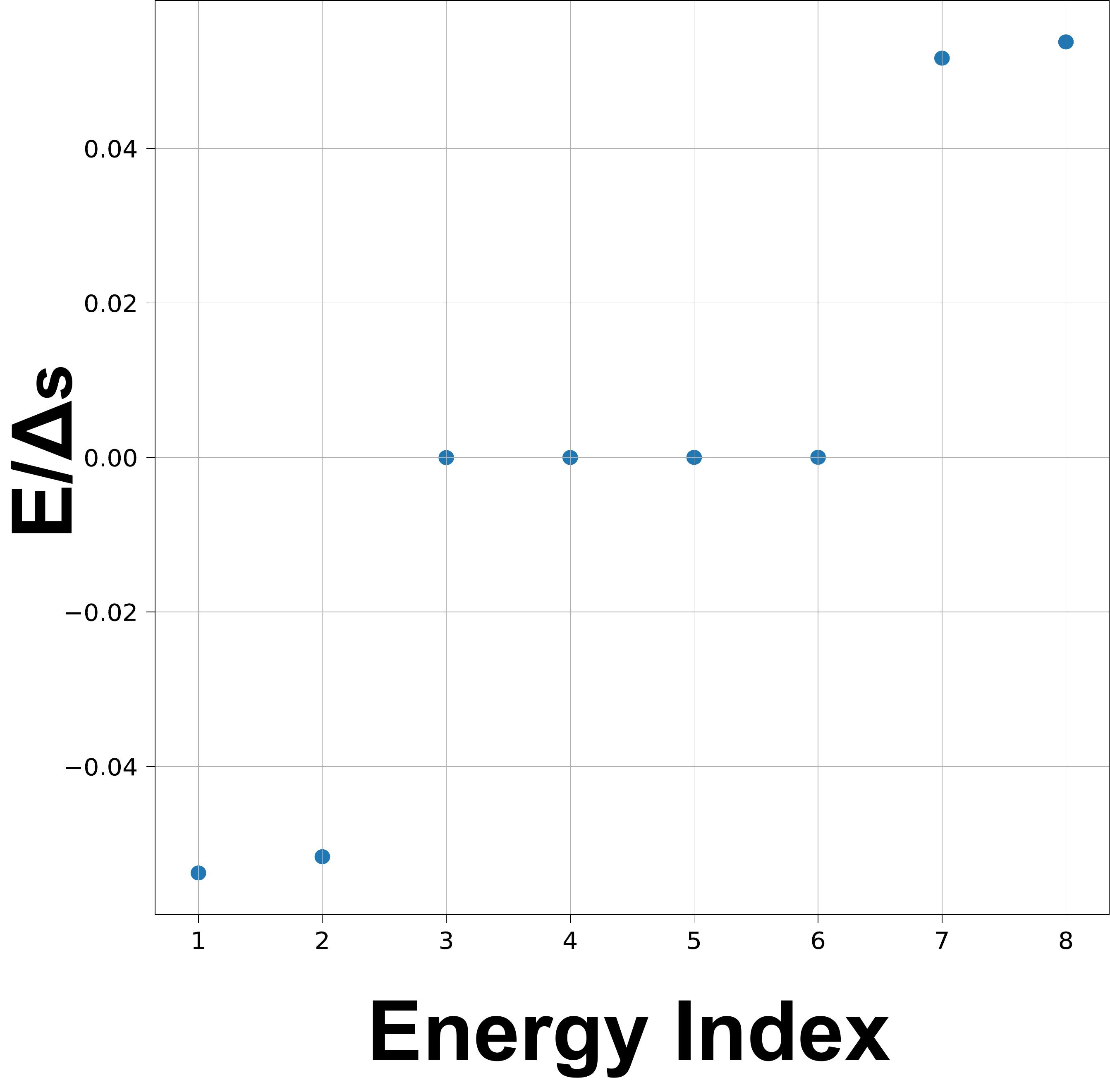}
    \caption{Numerical results in the 2D model with Zeeman coupling $mag=0.1$. In this calculation, the system involves $60\times60$ lattices. The parameters are: $\Delta_s=0.1$, $\Delta_t=0.25$, $\mu=-0.5$, $v_F=0.1$, $\xi=2$, $Q=2\pi/3$, $m'=0$ and $\epsilon'=0.25$. The calculation is performed with the kwant package~\cite{groth2014kwant}.}
    \label{zeeman}
\end{figure}

As shown in Fig.~\hyperref[triplet]{3(d)}, the system hosts four degenerate zero solutions in the spectrum.
This can be understood by decomposing the spin-triplet pairing $p_x\sigma_x+p_y\sigma_y$ into $p_x+ip_y$ and $p_x-ip_y$ chiral channels, each of which contributes one Majorana pair.
As $\Delta_t/\Delta_s$ decreases, the degeneracy is slightly lifted, which is due to the coupling between the two channels originating from the comparable $\Delta_s$. In general, in the absence of strong inter-channel mixing, this model can generate a doubled near-zero-mode spectrum in the vortex. Moreover, this degeneracy will not be broken by a small Zeeman field, as shown in Fig.~\ref{zeeman}. As a result, STM may still display a single robust zero-bias feature in realistic materials.

\section{Spectrum of 3D p-wave SC vortex}\label{appg}
\begin{figure}[h]
    \centering
    \includegraphics[width=0.3\linewidth]{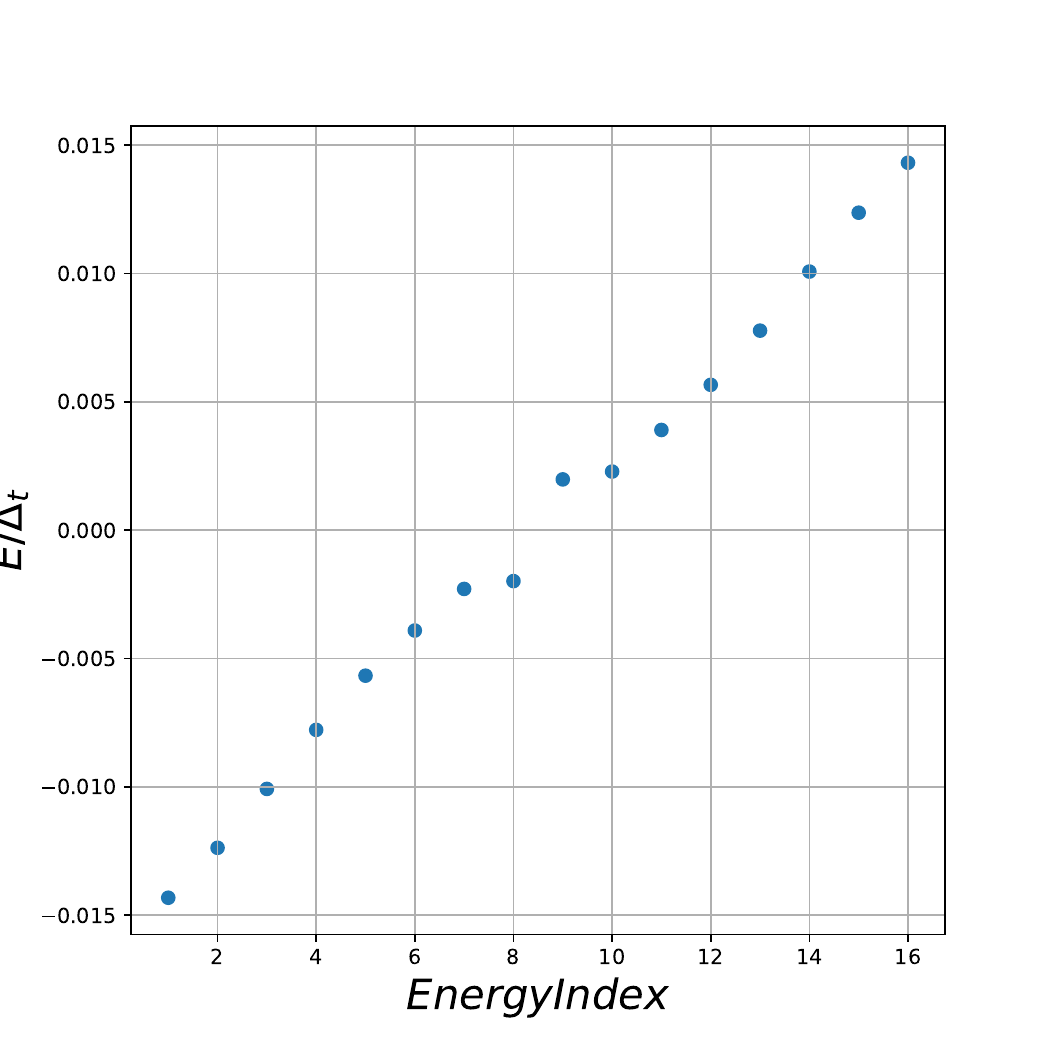}
    \caption{Numerical results in the 3D p-wave SC vortex. In our calculation, the system is a cylinder with radius $R=10$ and height $H=35$. Parameters are: $E_F=0.5$, $\Delta_s=0.0$, $\Delta_t=0.3$ $v_F=0.0$, $\xi=6$, $Q=\pi/2$, $m=0.14$ and $\epsilon=0.25$. The calculation is performed with the kwant package~\cite{groth2014kwant}.}
    \label{3Dpwave}
\end{figure}

We numerically calculate the spectrum of a 3D p-wave SC vortex. As stated in the maintext, in Fig.~\ref{3Dpwave}, the energy spacing is sufficiently smaller than $\Delta^2/E_F$, therefore the energy spectrum is nearly continuum.

\end{document}